\documentclass[aps%
,prb%
,showpacs%
,twocolumn%
,floats%
,amssymb%
]{revtex4}

\usepackage{graphicx}
\usepackage{amsmath}
\usepackage{amssymb}
\bibstyle{apsrev.bib}

\newcommand{\beq}{\begin{equation}}
\newcommand{\eeq}{\end{equation}}

\newcommand{\comment}[1]{}

\begin{document}
\title{Low-Temperature Phase Boundary of dilute Lattice Spin Glasses} 
\author{Stefan Boettcher and Emiliano Marchetti
} 

\affiliation{
Physics Department, Emory University, Atlanta, Georgia
30322, USA} 

\begin{abstract} 
The thermal-to-percolative crossover exponent $\phi$, well-known for
ferromagnetic systems, is studied extensively for Edwards-Anderson
spin glasses. The scaling of defect energies are determined at the
bond percolation threshold $p_c$, using an new algorithm. Simulations
extend to system sizes above $N=10^8$ in dimensions
$d=2,\ldots,7$. The results can be related to the behavior of the
transition temperature $T_g\sim(p-p_c)^{\phi}$ between the
paramagnetic and the glassy regime for $p\searrow p_c$. In three
dimensions, where our simulations predict $\phi=1.127(5)$, this
scaling form for $T_g$ provides a rare experimental test of
predictions arising from the equilibrium theory of low-temperature
spin glasses. For dimension near and above the upper critical
dimension, the results provide a new challenge to reconcile mean-field
theory with finite-dimensional properties.
\end{abstract} 
\maketitle

\comment{
Systems in this class possess a
low-$T$ glassy state with a transition temperature $T_g>0$. They are
characterized by a complex (free-)energy landscape in configuration
space with a hierarchy of valleys and barriers whose multi-modal
structure impedes the progression of any dynamics towards
equilibration, causing tantalizing phenomena, such as trapping and
jamming on intermediate time scales, and aging on long time scales. An
understanding of such systems is of paramount importance as these
phenomena are observed for a large class of materials as well as for
biological systems.\cite{Dagstuhl04}} 

The exploration of low-temperature properties of disordered systems
remains an important and challenging
problem.\cite{Young98,parisi-2007}  The paradigmatic model for such
phenomena is the Edwards-Anderson (EA) spin glass,\cite{F+H}
\begin{eqnarray}
H=-\textstyle{\sum_{<i,j>}}\,J_{i,j}\,x_i\,x_j,\quad(x_i=\pm1).
\label{Heq}
\end{eqnarray}
Disorder effects arise via quenched random bonds, $J_{i,j}$, mixing
ferro- and anti-ferromagnetic couplings between nearest-neighbor
spins, that lead to conflicting constraints and frustrated
variables. It is believed that an understanding of static and dynamic
features of EA may aid a description of the unifying principles
expressed in glassy materials.\cite{F+H} Most insights into
finite-dimensional systems has been gained through computational
approaches that elucidate low-$T$
properties.\cite{Binder86,Dagstuhl04,Boettcher04b}

Here, we extract the response induced through
defect-interfaces\cite{Southern77,Bray84} at $T=0$, created
by fixing the spins along the two faces of the open boundary in one
direction. Ground state energies
$E_0$ and $E_0'$ of an instance of size $N=L^d$ are determined that
differ by reversing all spins
on {\it one} of the faces. The distribution $P(\Delta E)$ of
interface energies $\Delta E=E_0'-E_0$ created by this perturbation of
scale $L$ on the boundary is obtained. The typical energy scale,
represented by the deviation $\sigma(\Delta E)$, grows as
\begin{eqnarray}
\sigma(\Delta E)\sim L^y.
\label{yeq}
\end{eqnarray}
This relation defines\cite{Southern77,Bray84,F+H} the stiffness exponent~$y$
 characterizing the defect energy, a
fundamental quantity assessing low-temperature fluctuations: a
positive value of $y$, as found\cite{Boettcher05d} in EA for
$d>d_l\approx5/2$, denotes the increase in the
energetic cost accompanying a growing number of variables perturbed
from their position in the ground state (i.~e., ``stiffness''). The
rise in strain for stronger disturbances signals the presence of an
ordered state. In turn, for systems with $y\leq0$ such order is
destabilized by fluctuations that spread unimpeded.

\begin{figure}[b!]
\vspace{1.27in}
\includegraphics{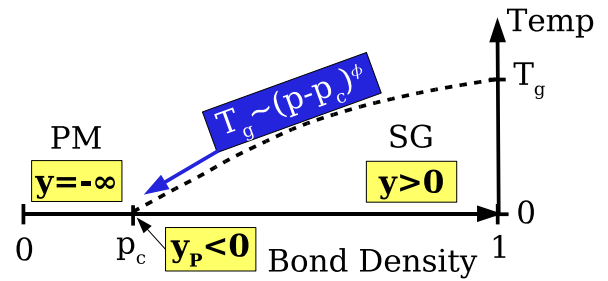}
\caption{(Color online) 
Phase diagram for bond-diluted spin glasses ($d>d_l$). In the
  spin-glass phase (SG) for $T<T_g$ and $p>p_c$, $y$ in
  Eq.~(\protect\ref{yeq}) is $>0$, while $y=y_P<0$ in Eq.~(\protect\ref{yPeq}) at $p=p_c$ and $T=T_g=0$. In the paramagnetic
  phase (PM) for $p<p_c$, defects decay
  exponentially for all $T$. The exponent $\phi$ in
  Eq.~(\protect\ref{Tgeq}) describes the boundary $T_g(p)$ for
  $p\searrow p_c$.
}
\label{PhaseDia}
\end{figure}

Instead of determining the interface scaling on a compact lattice
structure, we will focus here on the interface energy $\sigma(\Delta
E)$ on a {\it bond-diluted} lattice, in particular, at the percolation
threshold $p_c$, see Fig.~\ref{PhaseDia}.  Due to the tenuous fractal
nature of the percolating cluster at $p_c$, no long-range order can be
sustained, and\cite{Banavar87}
\begin{eqnarray}%
\sigma(\Delta E)_{L,p_c}\sim L^{y_P}\quad{\rm with}\quad y_P\leq0,
\label{yPeq}%
\end{eqnarray} 
i.~e., defects possess a vanishing interface energy. Interest in the exponent
$y_P$ stems from its relation to the ``thermal-percolative cross-over
exponent'' $\phi$ defined via\cite{Banavar87}
\begin{eqnarray}
T_g(p)\sim\left(p-p_c\right)^{\phi},\quad{\rm with}\quad \phi=-\nu
y_P
\label{Tgeq}
\end{eqnarray}
where $\nu$ is the correlation-length exponent associated with lattice
percolation,\cite{Stauffer94,Hughes96} $\xi\sim(p-p_c)^{-\nu}$. Of particular
experimental interest is the result for $d=3$, $y_P=-1.289(6)$, predicting $\phi=1.127(5)$
with $\nu=0.87436(46)$.\cite{Deng05} All results for $d=2,\ldots,7$
are listed in Tab.~\ref{alldata}.

\begin{table}
\caption{List of the parameters used and exponents found in our
  simulations for $d=2,\ldots,7$. $L_{\rm max}$ denotes the largest
  lattice size considered. We have used the bond-percolation
  thresholds $p_c$ from Ref.~\onlinecite{Lorenz98} for $d=3$ and
  Ref.~\onlinecite{Grassberger03} for $d\geq4$. The
  correlation-length exponents $\nu$ for percolation are from
  Ref.~\onlinecite{Deng05} in $d=3$ and from
  Ref.~\onlinecite{Hughes96} for $d\geq4$, where $\nu=1/2$ is exact
  above the upper critical dimension, $d\geq6$.}
\begin{tabular}{r|l|l|l|l|r}
\hline\hline
$d$& $p_c$        & $\nu$       & $y_P$      & $\phi=-\nu y_P$ & $L_{\rm max}$ \\
\hline
2  & 1/2          & 4/3         & -0.993(3)  &  1.323(4)       &  1000\\
3  & 0.2488126    & 0.87436(46) & -1.289(6)  &  1.127(5)       &   300\\
4  & 0.1601314    & 0.70(3)     & -1.574(6)  &  1.1(1)         &   100\\
5  & 0.118172     & 0.571(3)    & -1.84(2)   &  1.05(2)        &    35\\
6  & 0.0942019    & 0.5         & -2.01(4)   &  1.00(2)        &    25\\
7  & 0.0786752    & 0.5         & -2.28(6)   &  1.14(3)        &    15\\
\hline\hline
\end{tabular}
\label{alldata}
\end{table}

The exponent $\phi$ has been studied intensely numerically,
theoretically, and experimentally
\cite{Stephen77,Giri78,Southern79,Coniglio81,Aizenman87,Shapira94,Stauffer94,Munninghoff84}
for ferromagnetic systems some 30 years ago, and just recently was
discussed for quantum spins.\cite{Vojta07} But aside from its initial
treatment in Ref.~\onlinecite{Banavar87}, there are no investigations on
spin glasses. This is even more surprising, since this exponent
provides a non-trivial, experimentally testable prediction derived
from scaling arguments of the equilibrium theory at low
temperatures. Such tests are few as disordered materials by their very
nature fall out of equilibrium when entering the glassy state. The
phase boundary itself provides the perfect object for
such a study: It can be approached by theory from below and by
experiments from above where equilibration is possible.

There is reason to believe that the phase boundary in Eq.~(\ref{Tgeq})
and Fig.~\ref{PhaseDia} is experimentally accessible for certain
materials. Ref.~\onlinecite{Poon78} already provided highly accurate
results for the freezing temperature $T_M$ as a function of dilution
$x$ for a doped, cristalline glass, (La$_{1-x}$Gd$_x)_{80}$Au$_{20}$,
proposing a linear dependence, $T_M\sim x$. The
tabulated data is equally well fitted by Eq.~(\ref{Tgeq}) in that
regime. Ref.~\onlinecite{Beckman82} determined a phase diagram for
$({\rm Fe}_x{\rm Ni}_{1-x})_{75}{\rm P}_{16}{\rm B}_6{\rm Al}_4$, an
amorphous alloy, for a wide range of temperatures $T$ and
site-concentrations $x$ but did not discuss its near-linear behavior
at low $x$.  A similar phase diagram for the insulator
CdCr$_{2x}$In$_{2(1-x)}$S$_4$ can be found in Fig.~1.1a of
Ref.~\onlinecite{Vincent06}.  New experiments dedicated to the limit
$x\searrow x_c$ should provide results of sufficient accuracy to test
our prediction for $\phi$.

A match of computational prediction and experiment would lend
credibility to the EA model and its simplifying assumptions, such as
universality with respect to the details of the bond distribution
$P(J)$, an issue recently revisited by Ref.~\onlinecite{Katzgraber06}.
Our simulations, conducted here for Gaussian bonds, can be repeated
for any $P(J)$ of zero mean and unit variance, but would significantly
increase computational cost. Simply to demonstrate that theoretically
such universality exists, we have repeated our simulations with a
Lorentzian bond distribution argued for by Ref.~\onlinecite{Matho79},
and with powerlaw-distributed bonds $P(J)\propto |J|^{\alpha-1}$ for
$|J|\geq1$ at $\alpha=-1/2$. This comparison, presented below in
Fig.~\ref{compare}, clearly show reproducibility for a wide class of
$P(J)$. Ref.~\onlinecite{Banavar87} also considered power-law
distributions, but for $\alpha>0$ and $|J|<1$, to point out that the
response to perturbations is in principle non-universal at $p_c$:
Without long-range order, it is $y\leq0$ in Eq.~(\ref{yeq}) and energy
scales are \emph{not} diverging. Then, $y$ (and $\phi$) become
dependent on the details of $P(J)$ near $J=0$, and
Ref.~\onlinecite{Banavar87} finds an interesting change in behavior
for $\alpha<\alpha_c\approx0.75$. Such a diverging bond distribution
results from integrating the RKKY couplings over many weak bonds in
its far-distance tail. In realistic materials, such bonds are screened
out (see, for instance, Ref.~\onlinecite{Poon78}), and $P(J)$ is
bounded, justifying the use of Gaussian bonds, say.

Following the discussion in Refs.~\onlinecite{Banavar87,Bray87b}, for diluted
lattices at $p\to p_c$ we have to generalize the scaling relation for
the defect energy $\sigma(\Delta E)$ in Eq.~(\ref{yeq}) to
\begin{eqnarray}
\sigma(\Delta E)_{L,p}\sim {\cal Y}(p) L^y f\left(L/\xi(p)\right).
\label{sigmaeq}
\end{eqnarray}
Here, ${\cal Y}\sim(p-p_c)^t\sim\xi^{-t/\nu}$ is surface
tension and $\xi(p)\sim(p-p_c)^{-\nu}$ is the correlation length for
percolation. The scaling function $f$ is defined to be constant for
$L\gg\xi(p)\gg1$, where percolation (and hence, $\xi$)
plays no role and we regain Eq.~(\ref{yeq}) for $p>p_c$.

For $\xi\gg L\gg1$, Eq.~(\ref{sigmaeq}) requires $f(x)\sim
x^{\mu}$ for $x\to0$ to satisfy  $\sigma\to0$ with some
power of $L$, needed to cancel the $\xi$-dependence
at $p=p_c$. Thus, $\mu=-t/\nu$, and if we define $y_P=y+\mu=y-t/\nu$
to mark the $L$-dependence of $\sigma$ at $p=p_c$ as in
Eq.~(\ref{yPeq}), we get $t=\nu(y-y_P)$. Finally, at the cross-over
$\xi\sim L$, where the range $L$ of the excitations $\sigma(\Delta E)$
reaches the percolation length beyond which spin glass order ensues,
Eq.~(\ref{sigmaeq}) yields with $\phi$ from Eq.~(\ref{Tgeq}),
\begin{eqnarray}
\sigma(\Delta E)_{\xi(p),p}\sim\left(p-p_c\right)^t\xi(p)^yf(1)
\sim\left(p-p_c\right)^{\phi}.
\label{phieq}
\end{eqnarray}
Associating a temperature with this cross-over by $\sigma(\Delta
E)_{\xi(p),p}\sim T_g$ (for $T>T_g$, thermal fluctuation destroy
order), leads to Eq.~(\ref{Tgeq}), relating $p$ and $T_g$.

In our simulations we have used the method of bond
reductions described previously.\cite{MKpaper,Boettcher04c,Boettcher08a} A
set of rules is defined and applied recursively to trace out spins
assuming that $T=0$. These exact rules
apply to general Ising spin glass Hamiltonians as in Eq.~(\ref{Heq})
with {\it any} bond distribution $P(J)$, discrete or continuous, on
arbitrary sparse graphs, and lead to fewer
but more highly interconnected spins and renormalized bonds, see Ref.~\onlinecite{Boettcher08a}. Starting
from a Hamiltonian as in Eq.~(\ref{Heq}), in general, new terms are
generated by this procedure that have not been part of the Hamiltonian
before, such as multi-spin interaction. Although the
number of spins decreases one-by-one, the number of new terms grows
exponentially and the procedure usually becomes inefficient. Yet, near
$p_c$, we can apply a subset of these rules efficiently while
leaving the form of the 2-spin Hamiltonian in Eq.~(\ref{Heq}) invariant.

Our recursive set of rules is based on the following observations.
Near $p_c$, most spins have a low degree of interconnectivity; on
average, that degree fluctuates around unity in any dimension $d$. In
fact, many spins are entirely disconnected, do not contribute to the
Hamiltonian, and thus can be discarded. Degree-1 spins can always be
satisfied and are easily traced out, with their bond weight always (at
$T=0$) lowering the energy. Once all degree-1 spins have been
recursively traced out, any degree-2 spin can be reduced also by
replacing it by a new bond between its two neighbors and another
off-set to the global energy. Having reduced all degree-1 and -2
spins, there is even a ``star-triangle'' rule to reduce any degree-3
spins while only producing new 2-spin interactions between its
neighbors.\cite{Boettcher04c} Although this step in principle could
create a 3-spin interaction not present in the Hamiltonian in
Eq.~(\ref{Heq}), all such terms involving an odd number of spins
vanish due to the $Z_2$-symmetry of the Ising spins.

A new rule\cite{Boettcher08a} that proved particularly effective at $p_c$ focuses on
spins of arbitrary degree but with ``superbonds''.  A spin $x_i$ has a
superbond, if one bond's absolute weight dominates,
$|J_{i,k}|>\sum_{j\not=k}|J_{i,j}|$, all other bonds attached to
$x_i$. In the ground state ($T=0$), that bond is always satisfied and
its spin determined by its neighbor along that bond. This rule often
triggers new avalanches of further reductions with the simpler rules.

\begin{figure}[b!]
\vskip 2.05in 
\includegraphics{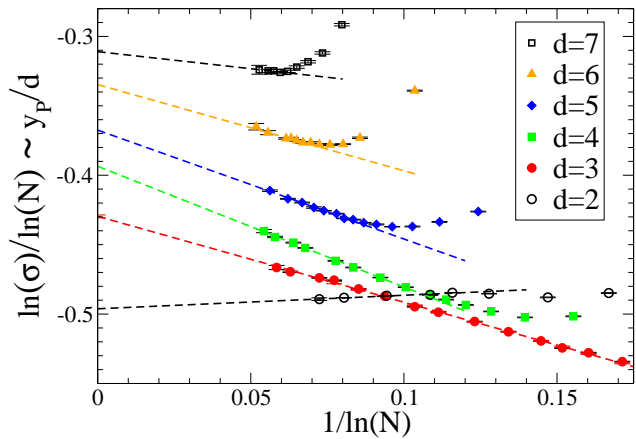}
\caption{(Color Online)
Plot of $\sigma\left(\Delta E\right)$ in Eq.~(\protect\ref{yPeq}) as a
function of system size $N=L^d$ in extrapolated form. Plotting
$\ln(\sigma)/\ln(N)$ vs. $1/\ln(N)$ and linearly extrapolating (dashed
lines), we extract the asymptotic values for $y_P/d$ in
Tab.~\protect\ref{alldata} at $1/\ln(N)=0$. Note that the fitted
asymptotic regime here corresponds to {\it orders of magnitude} of
scaling in a (less insightful) plot of $\ln\sigma$ vs. $\ln
N$. Note the increasing corrections to scaling for larger
$d$ before asymptotic behavior is obtained.
}
\label{fig:yP}
\end{figure}

Previously, we have applied these rules above $p_c$ to study the
defect energy within the spin glass state (SG), see
Fig.~\ref{PhaseDia}. Considering dilute lattices with $p>p_c$ but well
below $p=1$ allowed the study of larger lattice sizes $L$ for improved
scaling, and produced results\cite{Boettcher04c,Boettcher05d} in dimensions up to
$d=7$, unattainable with undiluted
lattices. For $p>p_c$ an optimization heuristic was essential to
approximate the ground state of the remainder graph, consisting of
highly interconnected spins that remain after all reduction rules have
been exhausted. In contrast, at $p=p_c$, these remained graphs are --
almost -- gone entirely. Thus, the attainable system sizes $N=L^d$ are
nearly unrestricted and have reached well above $N=10^8$ in our
simulations, mostly limited by the need to generate sufficient
statistics (i.~e., about $10^4$ instances for $N=25^6$ or $15^7$). Yet,
in $d=2$ and 3, the remainder graphs \emph{are} the limiting factor on
system sizes (at about $N=10^7$). Although remainders have less
than 100 spins, typically  well-approximated with a good
heuristic, we implemented costly exact methods\cite{Klotz94} to
optimize them. The {\it slightest} inaccuracy affected
the statistical averages, as defect energies $\Delta E$ are the
difference of two almost equal ground state energies $E_0$ and $E_0'$.
One technical problem in implementing our algorithm with such large
system sizes is posed by memory limitations. Instead of constructing
an entire lattice with $N$ spins, each with potentially $2d$ bonds,
before applying the reduction rules, we build up the $L^d$-spin
lattice as a sequence of $L$ hyper-planes of $L^{d-1}$ spins. During
the process, we keep the first and the most recently added plane
fixed, but already reduce recursively all spins in the intervening
planes as far as possible, before the next hyper-plane is added. This
process requires extensive bookkeeping and backtracking which can be
done fast while reducing memory use by $\sim1/L$.

\begin{figure}[b!]
\vskip 2.05in 
\includegraphics{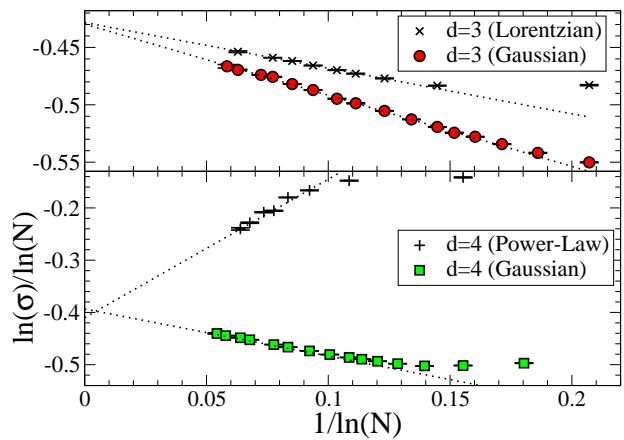}
\caption{(Color Online)
Comparison of the data for $d=3$ (top) and $d=4$ (bottom) for
different bond distributions $P(J)$, plotted in the same way as
Fig.~\ref{fig:yP}. There is little difference between the Gaussian and
the Lorentzian bonds even in $d=3$, as both are similarly smooth near
$J=0$. For our power-law bonds with vanishing support for $|J|<1$, our
methods are very inefficient. Yet, at least in $d=4$ those bonds
produce results far from but consistent with Gaussian bonds.
}
\label{compare}
\end{figure}

In Fig.~\ref{fig:yP} we present all data of our simulations for
$d=2,\ldots,7$ in an extrapolation plot. In Fig.~\ref{compare} we
compare the same data from $d=3$ ($d=4$) together with those from the
Lorentzian (power-law) bond distribution $P(J)$, as discussed above. Since our data
reaches above the upper critical dimension $d_u=6$ (of both,
percolation and spin glasses) and should approach mean-field behavior,
it is most natural to replace $L^{y_P}$ with
$N^{y_P/d}$ in Eq.~(\ref{yPeq}) and extrapolate for $y_P/d\sim
\ln(\sigma)/\ln(N)$. As Fig.~\ref{fig:yP} shows, aside from $d=2<d_l$,
the extrapolations for $d>d_l$ all seem to share common
characteristics and appear to vary smoothly with $d$. In particular,
we have pushed the simulations in $d=7$ to large enough $N$ to
conclude that there appears to be no drastic change in the scaling
behavior above $d_u$. Increasing corrections make it harder to reach
asymptotic scaling beyond $d=7$. Interestingly, all transients in
Fig.~\ref{fig:yP} themselves extrapolate to an intercept consistent
with $-1/2$, indicative of a higher-order correction term with a
$d$-independent exponent.

The result in $d=2$, where $T_g=0$, is very close to that
theoretically predicted in Ref.~\onlinecite{Banavar87},
$y_P\approx-0.99$, and could conceivably be $=-1$ exactly, as it is in
$d=1$ (where $p_c=1$). That would suggest that the spin glass on a
$d=2$ percolation cluster essentially consists of an extensive linear
backbone of bonds.

\begin{figure}[b!]
\vskip 2.125in 
\includegraphics{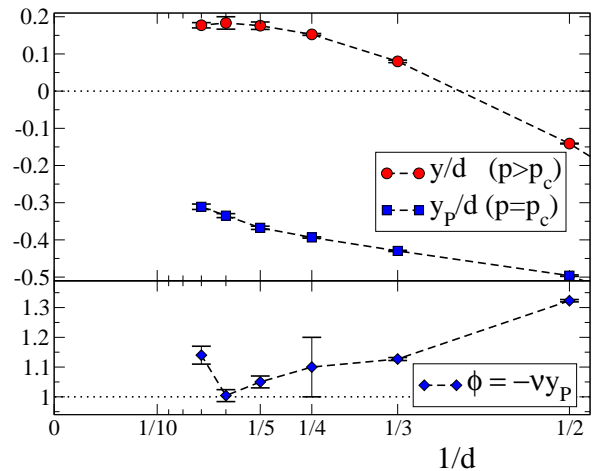}
\caption{(Color online)
Plot of the exponents $y_P/d$ (top) and $phi$ (bottom) in
Tab.~\protect\ref{alldata} as a function of $1/d$. For comparison,
also plotted (top) are the stiffness exponents $y/d$ inside the
spin-glass regime\cite{Boettcher05d} ($p>p_c$). Some of the
large error bars for $\phi=-\nu y_P$ originate with uncertainties in
$\nu$~\protect\cite{Hughes96}.}
\label{fig:highD}
\end{figure}

In Fig.~\ref{fig:highD}, we plotted $y_P/d$ and $\phi$ from
Tab.~\ref{alldata} vs. $1/d$ to explore the large-$d$ limit. This
extrapolation plot suggests a trend towards a vanishing value for
$y_P/d$ at $1/d=0$, i.~e., $y_P$ varies sub-linearly with $d$.  In
comparison, the data for the stiffness exponents $y/d$ inside the
spin-glass regime replotted from Ref.~\onlinecite{Boettcher05d} appears
consistent with the prediction\cite{Parisi07,Aspelmeier07} of
$y/d\sim1/6$. It would be difficult to suspect
a systematic bias in the apparent drift of the high-$d$ data points
for $y_P/d$, as the computations are exact. Yet, statistical errors
clearly become increasingly significant for larger $d$, see
Fig.~\ref{fig:yP}. It is not obvious how to directly obtain $y_P$ for
$d=\infty$, which may correspond to a (replica-symmetric) $T=T_g=0$
Viana-Bray model\cite{Viana85} at the Erd\"os-R\'enyi percolation
point. [Such a calculation has been undertaken for the
fully connected (replica-symmetry broken, $T<T_g$) SK
model.\cite{Aspelmeier03}] Finally, we note  a distinct minimum in
$\phi$, with $\phi_6\approx1$, exactly at the
upper critical dimension $d_u=6$, due to the product of increasing
$|y_P|$ and decreasing $\nu$.

This work has been supported by grant 0312510 from the Division of
Materials Research at the NSF and by the Emory
University Research Council. Thanks to S. Mertens for providing
computer time at Magdeburg University. We thank P. Nordblat
for helpful discussions.

\comment{
These commands specify that your bibliography should be created from entries in the file bibdata.bib. There are several bibliography styles that can be used with the eethesis document class; the style theunsrt.bst is the style used in the examples in this manual. The entries in this style are patterned after those in IEEE Transactions on Automatic Control. It lists the sources in the order they were cited. There is also a generic ieeetr.bst, which formats sources similar to many IEEE publications. If neither of these styles is suitable for your department, you might consider acm.bst or siam.bst which format your bibliography in the style of ACM and SIAM publications. Check the /usr/local/teTeX/local.texmf/bibtex/bst directory on the machine that you use to see if there are any other .bst files you can use.}
\bibliographystyle{apsrev}
\bibliography{/Users/stb/Boettcher}

\end{document}